\documentclass[12pt]{article}
\usepackage[top=1in, bottom=1in, left=1in, right=1in]{geometry}
\usepackage{authblk}
\usepackage{amsmath}
\usepackage{amssymb}
\usepackage{url}
\usepackage{subfigure}
\usepackage{natbib}
\usepackage[format=hang,labelfont=bf]{caption}
\usepackage[figuresright]{rotating}
\usepackage[english]{babel}
\usepackage{multirow}
\usepackage{booktabs}
\usepackage{hhline}
\usepackage{bm}
\usepackage{enumerate}
\usepackage{verbatim}
\usepackage{dsfont}
\usepackage[colorlinks=true,linkcolor=red,citecolor=blue]{hyperref}
\usepackage{threeparttable}
\usepackage{array}
\usepackage{makecell}
\usepackage{setspace}
\usepackage{lineno}
\usepackage{sectsty}
\sectionfont{\normalsize}
\subsectionfont{\small}
\usepackage{fancyhdr}
\newcommand{\sumi}{\sum_{i=1}^n}
\newcommand{\sumj}{\sum_{j=1}^{m_i}}
\newcommand{\sumk}{\sum_{k=1}^n}
\newcommand{\suml}{\sum_{l=1}^{m_k}}

\newcommand{\bfZ}{{\bf Z}}

\newcommand{\bfbeta}{\mbox{\boldmath $\beta$}}

\newcommand{\bfPsi}{\mbox{\boldmath $\Psi$}}

\newcommand{\bfzero}{{\bf 0}}
\newcommand{\bfI}{{\bf I}}
\newcommand{\bfV}{{\bf V}}

\newcommand{\bfD}{{\bf D}}
\newcommand{\bfC}{{\bf C}}
\newcommand{\bfc}{{\bf c}}
\newcommand{\bfS}{{\bf S}}

\newcommand{\bfU}{{\bf U}}

\newcommand{\bfalpha}{\mbox{\boldmath $\alpha$}}
\newcommand{\bfOmega}{\mbox{\boldmath $\Omega$}}
\newcommand{\ii}{{ij}}
\newcommand{\kl}{{kl}}

\newcommand{\intt}{\int_0^t}
\newcommand{\inta}{\int_0^\infty}

\newcommand{\Sz}{S^{(0)}}
\newcommand{\Sone}{\bfS^{(1)}}
\newcommand{\St}{\bfS^{(2)}}

\newcommand{\be}{\begin{eqnarray}}
\newcommand{\ee}{\end{eqnarray}}
\newcommand{\bee}{\begin{eqnarray*}}
\newcommand{\eee}{\end{eqnarray*}}
\newcommand{\bi}{\begin{enumerate}[(i)]}
\newcommand{\ei}{\end{enumerate}}

\title{Improving sandwich variance estimation for marginal Cox analysis of cluster randomized trials}

\author[*1,2]{Xueqi Wang}
\author[1,2]{Elizabeth L. Turner}
\author[3,4]{Fan Li}
\affil[1]{Department of Biostatistics and Bioinformatics, Duke University School of Medicine, Durham, NC, 27710, U.S.A.}
\affil[2]{Duke Global Health Institute, Duke University, Durham, NC, 27710, U.S.A.}
\affil[3]{Department of Biostatistics, Yale School of Public Health, New Haven, CT, 06511, U.S.A.}
\affil[4]{Center for Methods in Implementation and Prevention Science, Yale University, New Haven, CT, 06511, U.S.A.}
\renewcommand\Affilfont{\fontsize{9}{10.8}\selectfont}
{
    \makeatletter\scriptsize
    \renewcommand\AB@affilsepx{: \protect\Affilfont}
    \makeatother

    \affil[ ]{email}

    \makeatletter
    \renewcommand\AB@affilsepx{, \protect\Affilfont}
    \makeatother

    \affil[*]{xueqi.wang@duke.edu}
}
\date{}

\begin{document}
\maketitle
\thispagestyle{empty}
\label{firstpage}

\textbf{Abstract:}{ Cluster randomized trials (CRTs) frequently recruit a small number of clusters, therefore necessitating the application of small-sample corrections for valid inference. A recent systematic review indicated that CRTs reporting right-censored, time-to-event outcomes are not uncommon, and that the marginal Cox proportional hazards model is one of the common approaches used for primary analysis. While small-sample corrections have been studied under marginal models with continuous, binary and count outcomes, no prior research has been devoted to the development and evaluation of bias-corrected sandwich variance estimators when clustered time-to-event outcomes are analyzed by the marginal Cox model. To improve current practice, we propose $9$ bias-corrected sandwich variance estimators for the analysis of CRTs using the marginal Cox model, and report on a simulation study to evaluate their small-sample properties. Our results indicate that the optimal choice of bias-corrected sandwich variance estimator for CRTs with survival outcomes can depend on the variability of cluster sizes, and can also slightly differ whether it is evaluated according to relative bias or type I error rate. Finally, we illustrate the new variance estimators in a real-world CRT where the conclusion about intervention effectiveness differs depending on the use of small-sample bias corrections. The proposed sandwich variance estimators are implemented in an R package \textbf{CoxBcv}.}\\

\thispagestyle{empty}

\clearpage
\setcounter{page}{1}

\section{Introduction}\label{sec:intro}
Cluster randomized trials (CRTs) are frequently used to evaluate interventions in a range of settings from public health to education and social policy \citep{murray1998}. Reasons for choosing CRTs over individually randomized designs include advancing implementation convenience and minimizing contamination across experimental conditions \citep{turner2017a}, among others. A challenge in many CRT applications is that often a limited number of clusters may be available for randomization, even though the cluster size or the total sample size remains sufficiently large. For example, \citet{fiero2016statistical} performed a systematic review of 86 CRTs published between August $2013$ and July $2014$, and reported that the median number of clusters randomized was $24$. From a methodological review of around $300$ randomly selected CRTs published between $2000$ and $2008$, \citet{ivers2011impact} reported that the median number of clusters randomized was $21$. Frequently, including only a small number of clusters is due to logistical or resource constraints. For example, in a CRT evaluating a community based exercise program in over $65$ years old, only $12$ clusters were randomized by the reason of practical limitations \citep{munro2004cost}. Another example is a CRT accessing a palliative-care intervention to enable patients to spend more time at home, which only randomized $8$ clusters because of the limited resources available \citep{jordhoy2000palliative}. It has been documented that standard mixed-model or generalized estimating equations (GEE) analysis of CRTs with such small number of clusters (fewer than $30$) tend to inflate type I error rates, which necessitates the application of small-sample corrections to maintain valid inference \citep{murray2008design}.

A typical feature of CRTs is that outcome observations tend to be positively correlated within the same cluster. To account for such within-cluster correlations in the estimation of intervention effect in CRTs, one mainstream approach is the marginal regression model estimated by the GEE \citep{liang1986}. As elaborated by \citet{preisser2003}, this marginal approach bears a straightforward population-averaged interpretation and has the advantage of providing asymptotically valid inference through the so-called \emph{robust sandwich variance estimator} even when the within-cluster correlation structure is not correctly specified. However, a well-known limitation of the marginal analysis of CRTs is that the sandwich variance estimator can exhibit negative finite-sample bias when there is a small number of clusters (the typical rule of thumb for a small study is if it includes fewer than $30$ clusters), and several bias-corrected sandwich variance estimators have been developed to overcome this limitation; for example, by \citet{kauermann2001} (abbreviated as KC), \citet{mancl2001} (abbreviated as MD), \citet{fay2001} (abbreviated as FG), and \citet{morel2003} (abbreviated as MBN), which typically focused on generalized linear mean models. 

In general, these bias-corrected sandwich variance estimators hold strong promise for application to small CRTs, and there is a burgeoning literature on comparing their finite-sample performance in maintaining valid type I error rates with a continuous, binary or count outcome. To provide a few examples, \citet{Lu2007} examined the performance of the KC and MD variance estimators in pretest-posttest CRTs with a binary outcome, and reported that the KC variance estimator is often less biased with fewer than 30 clusters than the MD variance estimator. \cite{li2015} conducted comprehensive simulations to compare the KC, MD, FG, and MBN variance estimators in CRTs with a binary outcome simulated with both equal and unequal cluster sizes. They recommended to combine the Wald $t$-test with the KC variance estimator for controlling for the type I error rate under minor or moderate variation of the cluster sizes. With a larger degree of cluster size variation, however, the Wald $t$-test combined with the FG sandwich variance estimator had more robust small-sample performance. \citet{li2021sampleCount} made similar recommendations for analyzing small CRTs with count outcomes subject to possible right truncation. Furthermore, \citet{ford2017improved} proposed to use the average of the KC and MD standard errors for maintaining the nominal test size in small CRTs with continuous and binary outcomes, and \cite{Wang2016} provided a user-friendly R package for implementing these bias-corrected sandwich variance estimators with exponential family outcomes. The investigations of bias-corrected sandwich variance estimators have also been recently extended to more complex CRTs with multiple levels of clustering; see, for example, \citet{li2018}, \citet{ford2020maintaining}, \citet{thompson2021comparison}, and \citet{wang2021power}. 

Whereas the bias-corrected sandwich variance estimators have been extensively studied for GEE analysis of CRTs with continuous, binary and count outcomes, extensions and evaluations to censored time-to-event outcomes are relatively limited. \citet{caille2021methodological} reviewed 186 CRTs published from 2013 to 2018, and found that time-to-event outcomes are not uncommon but appropriate statistical methods are infrequently used. In particular, the marginal Cox proportional hazards model \citep{wei1989regression,lin1994cox} is a commonly used model for clustered right-censored time-to-event data. It models the effect of covariates on survival and hazard of an event occurrence, and targets the hazard ratio as a familiar effect measure. Assuming an independence working correlation structure, the marginal Cox model proceeds with estimating equations derived from the partial likelihood, and adjusts for the correlation through a robust sandwich variance estimator \citep{spiekerman1998marginal}. To date, there is limited guidance on appropriate bias-corrected sandwich variance estimators when there is a small number of clusters, for applications to CRTs as well as for more general clustered time-to-event analysis. To the best of our knowledge, \citet{fay2001} is the only study that performed a simulation evaluation of one specific bias-corrected sandwich variance estimator under the Cox model but not for clustered survival data. \textcolor{black}{More recently, \citet{chen2022clustered} studied the bias-corrected variance estimators for clustered restricted mean survival time regression models, \citet{blaha2022design} considered the bias-corrected variance estimators for CRTs with time-to-event outcomes under the additive hazards mixed model, and \citet{chen2022finite} compared different bias-corrected variance estimators for the marginal Fine-Gray regression model in CRTs with competing risks, but none of them studied the bias-corrected variance estimators for the Cox model analysis of CRTs.} To address this gap in statistical practice and improve finite-sample inference for small CRTs with time-to-event outcomes, we extend bias-corrected sandwich variance estimators developed for GEE analyses with non-censored outcomes to censored time-to-event outcomes. One of these corrected sandwich variance estimators exploits the bias in the martingale residual to inflate the original sandwich variance estimator, while others are based on multiplicative or additive adjustments. In Section \ref{sec:sim}, we further report the results of an extensive simulation study of these estimators under commonly seen CRT configurations, with equal and unequal cluster sizes, and discuss practical recommendations. In Section \ref{sec:app}, we provide an illustrative analysis of a pragmatic CRT, the Strategies and Opportunities to Stop Colorectal Cancer in Priority Populations (STOP CRC) trial, with $26$ clusters and a time-to-event outcome. Section \ref{sec:dis} concludes with a discussion and areas for future research. The proposed bias-corrected sandwich variance estimators for marginal Cox model are all implemented in the \textbf{CoxBcv} R package freely accessible on the Comprehensive R Archive Network (CRAN).

\section{Marginal Cox Proportional Hazards Model}\label{sec:gee}
Consider a parallel-arm CRT with $n$ clusters and $m_i$ as the cluster size for cluster $i$. For each individual $j\ (j = 1, \ldots, m_i)$ in cluster $i\ (i = 1, \ldots, n)$, we denote the observed data triplet as $\mathcal{O}_{ij}=(X_\ii, \Delta_\ii, \bfZ_\ii)$. In this data vector, we write $X_\ii = \text{min}\{T_\ii, C_\ii\}$ as the observed survival time, where $T_\ii$ is the underlying failure time for the event of interest, and $C_\ii$ is the censoring time; the event indicator $\Delta_\ii = 1$ if $X_\ii = T_\ii$ and $\Delta_\ii = 0$ if $X_\ii = C_\ii$. Finally, $\bfZ_\ii = (Z_{ij1}, \ldots, Z_{ijp})'$ is a $p \times 1$ vector of baseline covariates to be considered in the regression model. The covariate vector typically includes a cluster-level intervention status, and sometimes cluster-level and individual-level covariates of interest (in the former two cases, \textcolor{black}{elements of $\bfZ_\ii$ only depend on $i$ but not $j$}). Of note, the methods presented below can be easily generalized to time-dependent covariates $\bfZ_{ij}(t)$ with little modification, but we suppress the dependence on time where possible, to ease readability. In a parallel-arm CRT with time-to-event outcomes, the marginal Cox proportional hazards model is a common approach for estimating the population-averaged intervention effect, given by
\begin{align}\label{eq:ph}
    \lambda_\ii(t|\bfZ_\ii) = \lambda_0(t)\exp\left(\bfbeta'\bfZ_\ii\right),
\end{align}
where $\lambda_0(t)$ is an unspecified baseline hazard function and $\bfbeta$ is a $p \times 1$ vector of regression parameters. Frequently, the analysis of CRTs proceeds with only a cluster-level intervention indicator and no additional covariates, in which case $\bfZ_\ii$ is a scalar binary covariate and the scalar regression parameter is interpreted as the population-averaged hazard ratio. While our numerical evaluations focus on this simple case, we allow the presentation of the subsequent methodology to be more general with $p\geq 1$. 

\citet{wei1989regression} and \citet{lin1994cox} have discussed the details for estimating the population-averaged parameter $\bfbeta$ in the marginal Cox model and we provide a brief overview below. Based on an independence working correlation structure, an unbiased estimator for $\bfbeta$ in \eqref{eq:ph} can be obtained from the solution to the independence estimating equations
\begin{equation}\label{eq:ee}
\bfU(\bfbeta)=\sumi\bfU_i^\bullet(\bfbeta)= \sumi\sumj\Delta_\ii\left\{\bfZ_\ii - \frac{\Sone(\bfbeta; X_\ii)}{\Sz(\bfbeta; X_\ii)}\right\}=\bfzero,
\end{equation}
where $\bfS^{(r)}(\bfbeta; t) = \sumi\sumj Y_\ii(t)\exp\left(\bfbeta'\bfZ_\ii\right) \bfZ_\ii^{\otimes r}$ for $r=0, 1, 2$, $\bfc^{\otimes 0} = 1, \bfc^{\otimes 1} = \bfc, \bfc^{\otimes 2} = \bfc\bfc'$ for an arbitrary vector $\bfc$, and $Y_\ii(t) = I(X_\ii \geq t)$ is the at-risk process for each individual. This estimating equation is motivated based on the first-order condition of the partial likelihood formulation with independent and identically distributed survival data. With the estimated regression coefficients, $\widehat{\bfbeta}$, and denote $N_\ii(t) = I(X_\ii \leq t, \Delta_\ii = 1)$ as the counting process for the failure time, a consistent estimator for the cumulative baseline hazard function can be obtained with the Breslow-type estimator
\begin{gather*}
    \widehat{\Lambda}_0(t) = \sumi\sumj\intt\frac{dN_\ii(u)}{\sumk\suml Y_\kl(u)\exp\left(\bfbeta'\bfZ_{kl}\right)} = \sumi\sumj\intt\frac{dN_\ii(u)}{\Sz(\bfbeta; u)},
\end{gather*}
which implies that the baseline hazard can be estimated as $\widehat{\lambda}_0(t)dt = \sumi\sumj\Sz(\bfbeta; t)^{-1}dN_\ii(t)$.

In CRTs, although a valid point estimator for $\bfbeta$ can be obtained under the working independence assumption, the variance of $\widehat{\bfbeta}$ should be obtained from the robust sandwich variance estimator, which adjusts for the unknown within-cluster correlation structures and provides asymptotically correct uncertainty statements with clustered time-to-event data. The sandwich variance estimator has been studied extensively under generalized linear models and GEE with non-censored outcomes, and has been extended to the marginal Cox model by, for example, \citet{wei1989regression} and \citet{spiekerman1998marginal}. Define
\begin{align}\label{eq:u}
    \bfU_i(\bfbeta)=\sumj\bfU_\ii(\bfbeta)= \sumj\inta\left\{\bfZ_\ii - \frac{\Sone(\bfbeta; u)}{\Sz(\bfbeta; u)}\right\}dM_\ii(u), 
\end{align}
as the mean-zero martingale-score for each cluster, where 
$$M_\ii(t) = N_\ii(t) - \intt Y_\ii(u)\lambda_0(u)\exp(\bfbeta'\bfZ_{ij}) du$$ 
is the martingale with respect to the marginal filtration $\mathcal{F}_{t, ij}=\sigma\{N_{ij}(u),Y_{ij}(u),\bfZ_{ij},0 \leq u \leq t\}$ for each $i$, $j$, but not a martingale for the joint filtration due to intracluster correlations. Further define $\bfOmega_i(\bfbeta) = -{\partial\bfU_i(\bfbeta)}/{\partial\bfbeta}$ as the negative first-order derivative of $\bfU_i(\bfbeta)$ with respect to the regression parameter. Then the usual sandwich variance estimator (for example, implemented in the \textbf{coxph} function in \textbf{R}) is given by
\begin{align*}
    \widehat{\bfV}_s &= \widehat{\bfV}_m\left(\sumi\widehat\bfU_i\widehat\bfU_i'\right)\widehat{\bfV}_m,
\end{align*}
where
\begin{align*}
    \widehat{\bfV}_m &= \left(\sumi\widehat\bfOmega_i\right)^{-1} = \left( \sumi\sumj\inta\left\{\frac{\St(\widehat\bfbeta; u)}{\Sz(\widehat\bfbeta; u)} - \frac{\Sone(\widehat\bfbeta; u)\Sone(\widehat\bfbeta; u)'}{\Sz(\widehat\bfbeta; u)^2}\right\}dN_\ii(u) \right)^{-1}
\end{align*}
is the model-based variance estimator, and we write $\widehat\bfOmega_i = \bfOmega_i(\widehat\bfbeta)$, and $\widehat\bfU_i = \bfU_i(\widehat\bfbeta)$ for notational brevity. One typical feature of the standard sandwich variance estimator (we also refer to this estimator as the uncorrected sandwich variance estimator in subsequent text), $\widehat{\bfV}_s$, is that it is unbiased in large samples regardless of the correct specification of the working independent correlation assumption. However, when the number of clusters is small, as is more often the case in CRTs (frequently fewer than $30$), this default sandwich variance estimator tends to underestimate the variance, leading to inflated type I error rates and under-coverage \citep{li2022comparison}, which necessitates small-sample bias corrections to maintain valid statistical inference.

\section{Proposed Bias-Corrected Sandwich Variance Estimators}\label{sec:power}

\subsection{Bias correction based on modification of the martingale residual}\label{sec:MR}
\textcolor{black}{We first propose a bias correction based on the martingale residual (MR)}, similar to \citet{schaubel2005variance} for the analysis of clustered recurrent event data. Of note, $M_\ii(t)$ in Equation \eqref{eq:u} represents the mean-zero martingale, and we let $\widehat{M}(t)$ represent the martingale residual where the baseline hazard is estimated by the Breslow-type estimator and $\bfbeta$ is estimated by $\widehat{\bfbeta}$. We can rewrite the martingale as
\begin{align}\label{eq:res}
    M_\ii(t; \bfbeta) = \widehat{M}_\ii(t; \widehat\bfbeta) - \left\{\widehat{M}_\ii(t; \widehat\bfbeta) - \widehat{M}_\ii(t; \bfbeta)\right\} - \left\{\widehat{M}_\ii(t; \bfbeta) - M_\ii(t; \bfbeta)\right\},
\end{align}
where we define
\begin{align*}
    \widehat{M}_\ii(t; \bfbeta) &= N_\ii(t) - \intt Y_\ii(u)\widehat\lambda_0(u)\exp\left(\bfbeta'\bfZ_{ij}\right)du\\
    &= N_\ii(t) - \intt Y_\ii(u)\exp\left(\bfbeta'\bfZ_{ij}\right) \sumk\suml\frac{dN_\kl(u)}{\Sz(\bfbeta; u)}.
\end{align*}
In Web Appendix A, we consider a first-order Taylor Series expansion of $\widehat{M}_\ii(t; \widehat\bfbeta)$ around $\bfbeta$ such that Equation \eqref{eq:res} can be written as
\begin{align*}
    M_\ii(t; \bfbeta) = \widehat{M}_\ii(t; \widehat\bfbeta) + \widehat\bfD'_\ii(t; \bfbeta)\widehat\bfV_m\sumk\suml \bfU_\kl(\bfbeta) + 
    \intt Y_\ii(u)\exp\left(\bfbeta'\bfZ_\ii\right)\frac{dM(u)}{\Sz(\bfbeta; u)},
\end{align*}
where we define $M(t) = \sumi\sumj M_\ii(t)$ as the total sum of individual martingales and a gradient matrix 
\begin{align*}
    \widehat\bfD_\ii(t; \bfbeta) &= \intt \left\{\bfZ_\ii - \frac{\Sone(\bfbeta, u)}{\Sz(\bfbeta; u)}\right\}Y_\ii(u)\exp\left(\bfbeta'\bfZ_{ij}\right)\widehat\lambda_0(u)du.
\end{align*}

With a limited number of clusters, the aforementioned steps indicate that the martingale residual $\widehat{M}_\ii(t)$ can be biased for the martingale $M_\ii(t)$, therefore leading to potential bias in the sandwich variance estimator through $\widehat\bfU_i$. The bias of each individual martingale residual can be approximated by
\begin{align*}
    M_\ii(t; \bfbeta)-\widehat{M}_\ii(t; \widehat\bfbeta) \approx\widehat\bfD'_\ii(t; \widehat\bfbeta)\widehat\bfV_m\sumk\suml \bfU_\kl(\widehat\bfbeta) + 
    \intt Y_\ii(u)\exp\left(\widehat{\bfbeta}'\bfZ_\ii\right)\frac{d\widehat M(u)}{\Sz(\widehat\bfbeta; u)}.
\end{align*}
Based on this bias expression, we develop the following bias-corrected version of the estimated martingale-score $\widehat\bfU_i$ (additional details are provided in Web Appendix A):
\begin{align}
    \widehat\bfU_i^{BC} =& \left\{\bfI_{p} + \sumj\inta\left\{\bfZ_\ii - \frac{\Sone(\widehat\bfbeta; u)}{\Sz(\widehat\bfbeta; u)}\right\}d\widehat\bfD'_\ii(u; \widehat\bfbeta)\widehat\bfV_m\right\}\widehat\bfU_i \notag\\
     &+ \sumj\inta\left\{\bfZ_\ii - \frac{\Sone(\widehat\bfbeta; u)}{\Sz(\widehat\bfbeta; u)}\right\} Y_\ii(u)\exp\left(\widehat\bfbeta'\bfZ_\ii\right)\Sz(\widehat\bfbeta; u)^{-1}d\widehat M_{i\bullet}(u),\label{eq:bcU}
     \end{align}
where $\bfI_{p}$ is the $p\times p$ identity matrix, and $M_{i\bullet}(t) = \sumj M_\ii(t)$ is the sum of within-cluster martingales. The bias-corrected sandwich variance estimator for the marginal Cox model therefore takes the form of
\begin{align}\label{eq:varMR}
    \widehat\bfV_{MR} = \widehat{\bfV}_m\left\{\sumi\widehat\bfU_i^{BC}\left(\widehat\bfU_i^{BC}\right)'\right\}\widehat{\bfV}_m.
\end{align}

\subsection{Bias corrections based on methods for generalized estimating equations}\label{sec:mc}

We then generalize the class of multiplicative bias corrections developed for GEE to the sandwich variance estimator under the marginal Cox model. Following \citet{wang2021power}, the class of multiplicative bias corrections generally takes the form of $\widehat{\bfV}_m\widehat{\bfV}_0\widehat{\bfV}_m$ with
\begin{align}\label{eq:mc}
    \widehat{\bfV}_0 = \sumi\bfC_i\widehat\bfU_i\widehat\bfU_i'\bfC_i',
\end{align}
where $\bfC_i$ is the cluster-specific correction matrix to adjust for the negative bias in the sandwich variance estimator when the number of clusters is small. Specifically, to determine the form of $\bfC_i$,  we first notice that the score function can be written as $\bfU(\bfbeta)  = \sumi\bfU_i^\bullet(\bfbeta)+\sumi\bfU_i^\circ(\bfbeta)= \sumi\bfU_i(\bfbeta)$, where $\bfU_i^\bullet(\bfbeta)$ is defined in \eqref{eq:ee} and 
\begin{align}
\bfU_i^\circ(\bfbeta) &= -\sumj\sumk\suml\frac{\Delta_\kl Y_\ii(X_\kl)\exp\left(\bfbeta'\bfZ_{ij}\right)}{\Sz(\bfbeta, X_\kl)}\left\{\bfZ_\ii - \frac{\Sone(\bfbeta; X_\kl)}{\Sz(\bfbeta; X_\kl)}\right\},\notag
\end{align}
and further that $\sumi\bfU_i^\circ(\bfbeta) = \bfzero$ for all $\bfbeta$ \citep{lin1989robust}. This representation is not required for consistent estimation of the regression coefficients $\bfbeta$, but is merely used as a technical device to ensure that the estimating function from each cluster, $\bfU_i=\bfU_i(\bfbeta)$, are approximately mean zero and asymptotically independent across clusters (as opposed to $\bfU_i^\bullet(\bfbeta)$, which are not mean zero). This representation then allow us to employ the strategy of \citet{fay2001} and expand the estimating equations around $\widehat{\bfbeta}$ as $\bfU_i \approx \widehat\bfU_i - \widehat\bfOmega_i\left(\bfbeta-\widehat\bfbeta\right)$, where the hat notation indicates that the evaluation is at the estimator $\widehat\bfbeta$. By summing across all clusters and re-arranging terms, we obtain $\widehat\bfbeta - \bfbeta \approx \widehat\bfV_m\left(\sumi\bfU_i\right).$ Based on these intermediate results, we can approximate the covariance of the estimated cluster-specific score by (details are provided in Web Appendix B)
\begin{align}\label{eq:euu}
    \text{E}\left(\widehat\bfU_i\widehat\bfU_i'\right) \approx& \left(\bfI_p - \widehat\bfOmega_i\widehat\bfV_m\right)\bfPsi_i\left(\bfI_p - \widehat\bfOmega_i\widehat\bfV_m\right)' + \widehat\bfOmega_i\widehat\bfV_m\left(\sum_{j\neq i}\bfPsi_j\right)\widehat\bfV_m'\widehat\bfOmega_i',
\end{align}
where $\bfPsi_i = \text{Cov}(\bfU_i) = \text{E}\left(\bfU_i\bfU_i'\right)$ is the true covariance of the cluster-specific score, and the gradient matrix is estimated by
\textcolor{black}{
\begin{align*} 
     \widehat\bfOmega_i =& \sumj\inta\left\{\frac{\St(\widehat\bfbeta; u)}{\Sz(\widehat\bfbeta; u)} - \frac{\Sone(\widehat\bfbeta; u)\Sone(\widehat\bfbeta; u)'}{\Sz(\widehat\bfbeta; u)^2}\right\}dN_\ii(u)\\
     &- \sumj\inta\left\{\frac{\St(\widehat\bfbeta; u)}{\Sz(\widehat\bfbeta; u)} - \frac{\Sone(\widehat\bfbeta; u)\Sone(\widehat\bfbeta; u)'}{\Sz(\widehat\bfbeta; u)^2}\right\}Y_\ii(u)\text{exp}(\widehat\bfbeta'\bfZ_\ii) \sumk\suml\frac{dN_\kl(u)}{\Sz(\widehat\bfbeta; u)}\\
     &+ \sumj\inta\left\{\bfZ_\ii - \frac{\Sone(\widehat\bfbeta; u)}{\Sz(\widehat\bfbeta; u)}\right\} \bfZ_\ii' Y_\ii(u)\text{exp}(\widehat\bfbeta'\bfZ_\ii) \sumk\suml\frac{dN_\kl(u)}{\Sz(\widehat\bfbeta; u)}.
\end{align*}
}
We first assume that the working variance model is approximately within a scale of the true variance model as in \citet{kauermann2001} for GEE such that $\bfPsi_i \approx c\times \widehat\bfOmega_i$ (for example, this would be the case where the within-cluster correlation is relatively small, as is typically seen in applications to CRTs), and obtain from \eqref{eq:euu}
\begin{align*}
    \text{E}\left(\widehat\bfU_i\widehat\bfU_i'\right) &\approx\left(\bfI_p - \widehat\bfOmega_i\widehat\bfV_m\right)\bfPsi_i \approx \bfPsi_i\left(\bfI_p - \widehat\bfOmega_i\widehat\bfV_m\right)'.
\end{align*}
This motivates the choice of the correction matrix $\bfC_i = \left(\bfI_p - \widehat\bfOmega_i\widehat\bfV_m\right)^{-1/2}$ in Equation \eqref{eq:mc}, and we denote the resulted bias-corrected sandwich variance estimator by
\begin{align}\label{eq:varKC}
\widehat\bfV_{KC}=\widehat{\bfV}_m\left\{\sumi \left(\bfI_p - \widehat\bfOmega_i\widehat\bfV_m\right)^{-1/2}\widehat\bfU_i\widehat\bfU_i'\left(\bfI_p - \widehat\bfV_m\widehat\bfOmega_i\right)^{-1/2}\right\}\widehat{\bfV}_m.
\end{align}
Alternatively, the choice of the correction matrix can be analogous to \citet{fay2001}, by setting $\bfC_i = \text{diag}\left\{\left(1-\text{min}\left\{r, [\widehat\bfOmega_i\widehat\bfV_m]_{jj}\right\}\right)^{-1/2}\right\}$ in Equation \eqref{eq:mc}, where $r < 1$ is a user-defined constant. This choice of correction matrix corresponds to the the second multiplicative bias correction, resulting in the bias-correction sandwich variance estimator
\begin{align}\label{eq:varFG}
%\resizebox{0.94\hsize}{!}{
\widehat\bfV_{FG}=
\widehat{\bfV}_m &\left[\sumi\text{diag}\left\{\left(1-\text{min}\left\{r, [\widehat\bfOmega_i\widehat\bfV_m]_{jj}\right\}\right)^{-1/2}\right\}\widehat\bfU_i\widehat\bfU_i'\right.\notag\\
&\quad\quad \left. \vphantom{\sumi} \times\text{diag}\left\{\left(1-\text{min}\left\{r, [\widehat\bfOmega_i\widehat\bfV_m]_{jj}\right\}\right)^{-1/2}\right\}\right]\widehat{\bfV}_m.
\end{align}
Particularly, we choose $r = 0.75$ as default to ensure the maximum diagonal element of the correction matrix is bounded above by 2 and avoid over-correction \citep{fay2001}. Oftentimes, primary analysis of CRTs proceeds without covariates; in this case $\bfZ_{ij}$ only includes a cluster-level intervention indicator ($p=1$), so $\widehat\bfV_{FG}$ and $\widehat\bfV_{KC}$ would be numerically identical if $\widehat\bfOmega_i\widehat\bfV_m$ does not exceed the constant $r$, even though their values are generally different otherwise. The above derivations are also compatible with the insight in \citet{ziegler2011generalized} that the KC sandwich variance can be derived as a modified version of the FG sandwich variance, and the numerical evidence in \citet{wang2021power} that these two often have similar finite-sample operating characteristics for GEE analyses of non-censored outcomes. A final multiplicative bias-correction is analogous to \citet{mancl2001}, and assumes that the last term of \eqref{eq:euu} is negligible, which then gives
\begin{align*}
    \bfPsi_i &\approx \left(\bfI_p - \widehat\bfOmega_i\widehat\bfV_m\right)^{-1} \widehat\bfU_i\widehat\bfU_i' \left(\bfI_p - \widehat\bfV_m\widehat\bfOmega_i\right)^{-1}.
\end{align*}
This motivates the choice of correction matrix as $\bfC_i = \left(\bfI_p - \widehat\bfOmega_i\widehat\bfV_m\right)^{-1}$ in Equation \eqref{eq:mc}, resulting in a bias-corrected sandwich variance estimator denoted by
\begin{align}\label{eq:varMD}
\widehat\bfV_{MD}=\widehat{\bfV}_m\left\{\sumi \left(\bfI_p - \widehat\bfOmega_i\widehat\bfV_m\right)^{-1}\widehat\bfU_i\widehat\bfU_i'\left(\bfI_p - \widehat\bfV_m\widehat\bfOmega_i\right)^{-1}\right\}\widehat{\bfV}_m.
\end{align} Generally, the diagonal element of $\left(\bfI_p - \widehat\bfOmega_i\widehat\bfV_m\right)$ takes values between $0$ and $1$, and therefore $\widehat\bfV_{MD}$ often leads to larger variance estimates than $\widehat\bfV_{KC}$ \citep{Lu2007}. 

Finally, we consider an additive bias correction, analogous to \cite{morel2003} for GEE analyses of non-censored outcomes, defined as
\begin{align}\label{eq:varMBN}
    \widehat\bfV_{MBN} &= \left(\frac{\sumi m_i-1}{\sumi m_i-p}\times\frac{n}{n-1}\right)\widehat\bfV_s + \text{min}\left(0.5, \frac{p}{n-p}\right)\widehat\phi\widehat\bfV_m,
\end{align}
where
\begin{align*}
\widehat\phi = \text{max}\left\{1, \left(\frac{\sumi m_i-1}{\sumi m_i-p}\times\frac{n}{n-1}\right)\times\text{trace}\left[\widehat\bfV_m\left(\sumi\widehat\bfU_i\widehat\bfU_i'\right)\right]/p\right\}.
\end{align*}
Here, the quantity of $\text{trace}\left[\widehat\bfV_m\left(\sumi\widehat\bfU_i\widehat\bfU_i'\right)\right]$ in $\widehat{\phi}$ represents the sum of eigenvalues of the matrix $\widehat\bfV_m\left(\sumi\widehat\bfU_i\widehat\bfU_i'\right)$, and has been referred to as the generalized design effect \citep{rao1981analysis}. \citet{morel2003} showed that this additive bias-correction works well in small samples for GEE with commonly employed working correlation models. Compared to the class of multiplicative bias-corrections, one potential advantage of the additive bias correction is that it ensures a positive-definite covariance matrix \citep{morel2003}. With any of the multiplicative or additive bias-corrections, as the number of clusters $n$ increases, the bias-corrected sandwich variance estimator converges to the same probability limit as the uncorrected sandwich variance estimator, as can be seen from, for example, the fact that $\bfC_i\stackrel{p}{\rightarrow} \bfI_p$. Therefore, there is no harm asymptotically in employing the bias-corrected sandwich variance estimators for inference in small CRTs, and more generally for clustered time-to-event outcomes.

\subsection{Hybridizing bias corrections}\label{sec:hybrid}
Because the time-to-event analysis of CRTs concerns incompletely observed outcomes, the challenges in small-sample inference can be magnified due to right censoring, and it is likely that each bias-correction technique alone studied in Section \ref{sec:MR} and Section \ref{sec:mc} may be insufficient when implemented alone. An intuitive further adjustment is to hybridize the martingale residual-based correction with either one of the multiplicative or additive bias-corrections studied in Section \ref{sec:mc} to further remove the finite-sample bias with a small number of clusters. Specifically, we replace $\widehat\bfU_i$ in Equations \eqref{eq:varKC}, \eqref{eq:varFG}, \eqref{eq:varMD} and \eqref{eq:varMBN} with $\widehat\bfU_i^{BC}$ derived in \eqref{eq:bcU}. This defines four additional hybrid bias-corrected sandwich variance estimators, which we denote by  $\widehat\bfV_{KCMR}, \widehat\bfV_{FGMR}, \widehat\bfV_{MDMR}$ and $\widehat\bfV_{MBNMR}$. To better keep track of the proposed bias-corrected sandwich variance estimators, a brief summary of different variance estimators is provided in Table \ref{tb:Table1}.

\begin{table}[htbp]
\caption{A brief summary of uncorrected and $9$ bias-corrected sandwich variance estimators for the marginal Cox proportional hazards model for analyzing clustered data.}\label{tb:Table1}
\centering
\resizebox{\textwidth}{!}{%
\begin{threeparttable}
\begin{tabular}{lccc}
\toprule
%\midrule
Variance Estimator & Label & Formula & Feature \\
\midrule\smallskip
$\widehat{\bfV}_s$ & ROB & \multirow{4}[14]{*}{$\widehat{\bfV}_m\left(\sumi\bfC_i\widehat\bfU_i\widehat\bfU_i'\bfC_i'\right)\widehat{\bfV}_m$} & $\bfC_i = \bfI_p$ \\\cmidrule{4-4}
$\widehat\bfV_{KC}$ & KC & & $\bfC_i = \left(\bfI_p - \widehat\bfOmega_i\widehat\bfV_m\right)^{-1/2}$ \\\cmidrule{4-4}
$\widehat\bfV_{FG}$ & FG & & $\bfC_i = \text{diag}\left\{\left(1-\text{min}\left\{r, [\widehat\bfOmega_i\widehat\bfV_m]_{jj}\right\}\right)^{-1/2}\right\}$ \\\cmidrule{4-4}
$\widehat\bfV_{MD}$ & MD & & $\bfC_i = \left(\bfI_p - \widehat\bfOmega_i\widehat\bfV_m\right)^{-1}$ \\\midrule
$\widehat\bfV_{MBN}$ & MBN & $c_1\widehat\bfV_s + c_2\widehat\phi\widehat\bfV_m$ & $c_1, c_2, \widehat\phi$ defined in Equations \eqref{eq:varMBN} \\\midrule
$\widehat\bfV_{MR}$ & MR & \multirow{4}[14]{*}{$\widehat{\bfV}_m\left(\sumi\bfC_i\widehat\bfU_i^{BC}\left[\widehat\bfU_i^{BC}\right]'\bfC_i'\right)\widehat{\bfV}_m$} & $\bfC_i = \bfI_p$ \\\cmidrule{4-4}
$\widehat\bfV_{KCMR}$ & KCMR & & $\bfC_i = \left(\bfI_p - \widehat\bfOmega_i\widehat\bfV_m\right)^{-1/2}$ \\\cmidrule{4-4}
$\widehat\bfV_{FGMR}$ & FGMR & & $\bfC_i = \text{diag}\left\{\left(1-\text{min}\left\{r, [\widehat\bfOmega_i\widehat\bfV_m]_{jj}\right\}\right)^{-1/2}\right\}$ \\\cmidrule{4-4}
$\widehat\bfV_{MDMR}$ & MDMR & & $\bfC_i = \left(\bfI_p - \widehat\bfOmega_i\widehat\bfV_m\right)^{-1}$ \\\midrule
$\widehat\bfV_{MBNMR}$ & MBNMR & $c_1\widehat\bfV_{MR} + c_2\widehat\phi\widehat\bfV_m$ & \makecell{$c_1, c_2, \widehat\phi$ defined in Equations \eqref{eq:varMBN} with \\ $\widehat\bfU_i$ replaced by $\widehat\bfU_i^{BC}$} \\
\bottomrule
\end{tabular}%\smallskip
%\begin{tablenotes}\scriptsize
%\item[a]
%\end{tablenotes}
\end{threeparttable}
}
\end{table}

\section{Numerical Studies}\label{sec:sim}
Based on the marginal Cox analysis of CRTs, we are only aware of \citet{zhong2015sample}, who conducted a simulation study to examine sample size considerations. While their study is informative insofar as study design considerations, the smallest number of clusters included was $71$, which is substantially larger than what is generally thought realistic in biomedical and public health studies involving CRTs; their scenarios are also precisely the cases where the uncorrected robust variance estimator has negligible bias. To generate practical recommendations for common practice, we designed an extensive simulation study to evaluate the performance of the $9$ bias-corrected sandwich variance estimators compared to the uncorrected sandwich variance estimator, with a small or moderate number of clusters under the marginal Cox model.

\subsection{Study description}\label{sec:msd}
We considered a two-arm CRT with equal allocation, and the only covariate in model \eqref{eq:ph} was a binary cluster-level intervention indicator (commonly referred to as the unadjusted analysis, which is the standard primary analysis in most CRTs), with $Z_i = 1$ indicating the intervention arm and $Z_i = 0$ indicating the control arm. We were interested in testing the null hypothesis of no intervention effect $H_0: \beta = 0$, using a two-sided Wald $t$-test with $n-1$ degrees of freedom (since the model only includes an intervention indicator variable); the $t$-test was adopted because it generally performs better in finite samples than the $z$-test \citep{li2015,wang2021power}. Our data-generating process and parameter specification follow the approach in \citet{zhong2015sample}. We assumed a proportional hazards model as in \eqref{eq:ph}, where the cumulative baseline hazard was specified by a Weibull distribution with $\Lambda_0(t; \bfalpha) = \intt\lambda_0(s; \bfalpha)ds = (\lambda_0t)^\kappa$ and $\bfalpha = (\lambda_0, \kappa)'$. We fixed the administrative censoring time as $C^\dagger = 1$, so that all of the observations were in $(0, C^\dagger]$. The parameter $\lambda_0$ was chosen as the solution to $\text{P}(T_\ii > C^\dagger|Z_i = 0) = p_a$ to give the desired administrative censoring rate for the control group. In particular, we selected $p_a = 0.2$. A random censoring time for individual $j$ in cluster $i$ was denoted by $C^*_\ii$ and assumed to be exponentially distributed with rate $\rho$, and we assumed that censoring was independent within each cluster. Therefore, the true right-censoring time was $C_\ii = \text{min}\{C^*_\ii, C^\dagger\}$. The parameter $\rho$ was chosen as the solution to $\text{P}(T_\ii > C_\ii|Z_i = 0) = p_0$ to give the desired net censoring rate in the control arm, where we considered $p_0 = 0.2$ for the case of strictly administrative censoring, due to the fact of $p_0 = p_a$ in this case, and $p_0 = 0.5$ for the case of $20\%$ administrative and $30\%$ random censoring. 

Correlated failure times in each cluster were generated from the Clayton copula \citep{clayton1985multivariate} and also followed the data-generating process in \citet{cai2000permutation}. In general, based on Clayton copula specification, the joint survival distribution for $m_i \ (m_i \geq 2)$ correlated observations $(T_{i1}, \ldots, T_{im_i})$ in a cluster is
\begin{align*}
    S_i(t_{i1}, \ldots, t_{im_i}) = \text{P}(T_{i1}>t_{i1}, \ldots, T_{im_i}>t_{im_i}) = \left\{\sum_{j=1}^{m_i}S_{ij}(t_{ij})^{-1/\theta}-(m_i-1)\right\}^{-\theta},
\end{align*}
and the conditional cumulative distribution function for $T_{i1}, \ldots, T_{ih} \ (h = 1, \ldots, m_i)$ is
\begin{align*}
    F_{ih}\left(t_{ih} \mid t_{i1}, \ldots, t_{i,h-1}\right) = 1-&\left[\left\{\sum_{j=1}^{h} S_{ij}\left(t_{ij}\right)^{-1 / \theta}-(h-1)\right\}^{-\theta-h+1} \right.\\
    &\quad \left. \times \left\{\sum_{j=1}^{h-1} S_{ij}\left(t_{ij}\right)^{-1 / \theta}-(h-2)\right\}^{\theta+h-1}\right],
\end{align*}
where $S_{ij}(t) = \text{P}(T_{ij}>t)$ is the marginal survival function for $T_{ij}$, and $\theta$ characterizes the degree of association between failure times within a cluster. As a cumulative distribution function, $F_{ih}\left(t_{ih} \mid t_{i1}, \ldots, t_{i,h-1}\right) \sim \text{Uniform}(0, 1)$, so we can generate $m_i$ independent $\text{Uniform}(0, 1)$ variates, denoted as $(u_{i1}, \ldots, u_{im_i})$, i.e., $u_{i1} = F_{i1}(t_1)$ and $u_{ih} = F_{ih}\left(t_{ih} \mid t_{i1}, \ldots, t_{i,h-1}\right)$ for $h = 2, \ldots, m_i$. Then solving for $t_{i1}$ and $t_{ih} \ (h = 2, \ldots, m_i)$ gives $t_{i1} = S_{i1}^{-1}(1-u_{i1})$ and
\begin{align*}
    t_{ih} = S_{ih}^{-1}\left(\left[(h-1)-\sum_{j=1}^{h-1} a_{ij}+\left\{\sum_{j=1}^{h-1} a_{ij}-(h-2)\right\}\left(1-u_{ih}\right)^{-(\theta+h-1)^{-1}}\right]^{-\theta}\right)
\end{align*}
for $h = 2, \ldots, m_i$, where
\begin{align*}
    a_{il} = S_{il}\left(t_{il}\right)^{-1/\theta} = (l-1)-\sum_{j=1}^{l-1} a_{ij}+\left\{\sum_{j=1}^{l-1} a_{ij}-(l-2)\right\}\left(1-u_l\right)^{-(\theta+l-1)^{-1}}
\end{align*}
for $l = 1, \ldots, h-1$.

With the assumption that the cumulative baseline hazard was specified by a Weibull distribution, the marginal survival function for $T_\ii \ (i = 1, \ldots, n; j = 1, \ldots, m_i)$ was $S_\ii(t) = \text{exp}\left\{-\left(\lambda_0 t\right)^\kappa\text{exp}\left(Z_i\beta\right)\right\}$, and its inverse function was $S_\ii^{-1}(\eta) = \left\{-\text{log}(\eta)\times\text{exp}(-Z_i\beta)\right\}^{1/\kappa}/\lambda_0$. Also, we varied the copula parameter $\theta$ such that the Kendall's $\tau=1/(2\theta+1)$ is $0.01, 0.05, 0.1$, and $0.25$ for small, mild, moderate, and large within-cluster associations, respectively. Therefore, correlated failure times $(t_{i1}, \ldots, t_{im_i})$ in cluster $i \ (i = 1, \ldots, n)$ were generated through the following procedure. (i) Generate $m_i$ independent $\text{Uniform}(0, 1)$ variates $(u_{i1}, \ldots, u_{im_i})$. (ii) Calculate $t_{i1} = \left\{-\text{log}(1-u_{i1})\times\text{exp}(-Z_i\beta)\right\}^{1/\kappa}/\lambda_0$, and $a_{i1} = (1-u_{i1})^{-1/\theta}$. (iii) For $h = 2, \ldots, m_i$, calculate $\omega_{ih} = (h-1)-\sum_{j=1}^{h-1} a_{ij}+\left\{\sum_{j=1}^{h-1} a_{ij}-(h-2)\right\}$ $\times\left(1-u_{ih}\right)^{-(\theta+h-1)^{-1}}$, where $a_{il} = (l-1)-\sum_{j=1}^{l-1} a_{ij}+\left\{\sum_{j=1}^{l-1} a_{ij}-(l-2)\right\}\left(1-u_{il}\right)^{-(\theta+l-1)^{-1}}$ for $l = 1, \ldots, h-1$; then calculate $t_{ih} = \left\{\theta\text{log}(\omega_{ih})\times\text{exp}(-Z_i\beta)\right\}^{1/\kappa}/\lambda_0$.

We set the marginal mean parameter $\beta = 0$ for assessing the empirical type I error rate. The nominal type I error rate was fixed at $5\%$. We chose the number of clusters $n \in \{6, 10, 20, 30\}$, and cluster sizes $m_i \ (i = 1, \ldots, n)$ were generated from a gamma distribution with mean equal to $\overline{m} \in \{20, 50, 100\}$ and coefficient of variation (CV) ranging from $0$ to $1$ by an increment of $0.1$. The smallest cluster size will be truncated at $2$ to ensure numerical stability. For each scenario, we generated $5000$ data replications and fit the marginal Cox model for each replicate. We considered $10$ different variance estimators for the intervention effect: the uncorrected robust sandwich variance estimator $\widehat{V}_s$ (denoted as ROB), and $9$ bias-corrected sandwich variance estimators summarized in Table \ref{tb:Table1}. The results of interest were the percent relative bias of the variance estimators, and the empirical type I error rate under the null. Specifically, the percent relative bias of the evaluated variance estimator, indexed as $q$, was calculated as $\left\{\sum_{r=1}^{5000}(\widehat{V}_q)_r/5000 - \text{Var}_{MC}(\widehat{\beta})\right\}/\text{Var}_{MC}(\widehat{\beta}) \times 100$, where $(\widehat{V}_q)_r$ was $\widehat{V}_q$ from the $r$th simulated data replication, and $\text{Var}_{MC}(\widehat{\beta}) = \sum_{r=1}^{5000}(\widehat{\beta} - \sum_{r=1}^{5000}\widehat{\beta}/5000)^2/4999$.  In addition, given the nominal type I error rate was $5\%$, we considered an empirical type I error rate from $4.4\%$ to $5.6\%$ as close to nominal by the margin of error from a binomial model with $5000$ replications \citep{morris2019using}.

\subsection{Study results}\label{sec:sr}
\begin{figure}[htbp]
\centering
\includegraphics[scale=0.51]{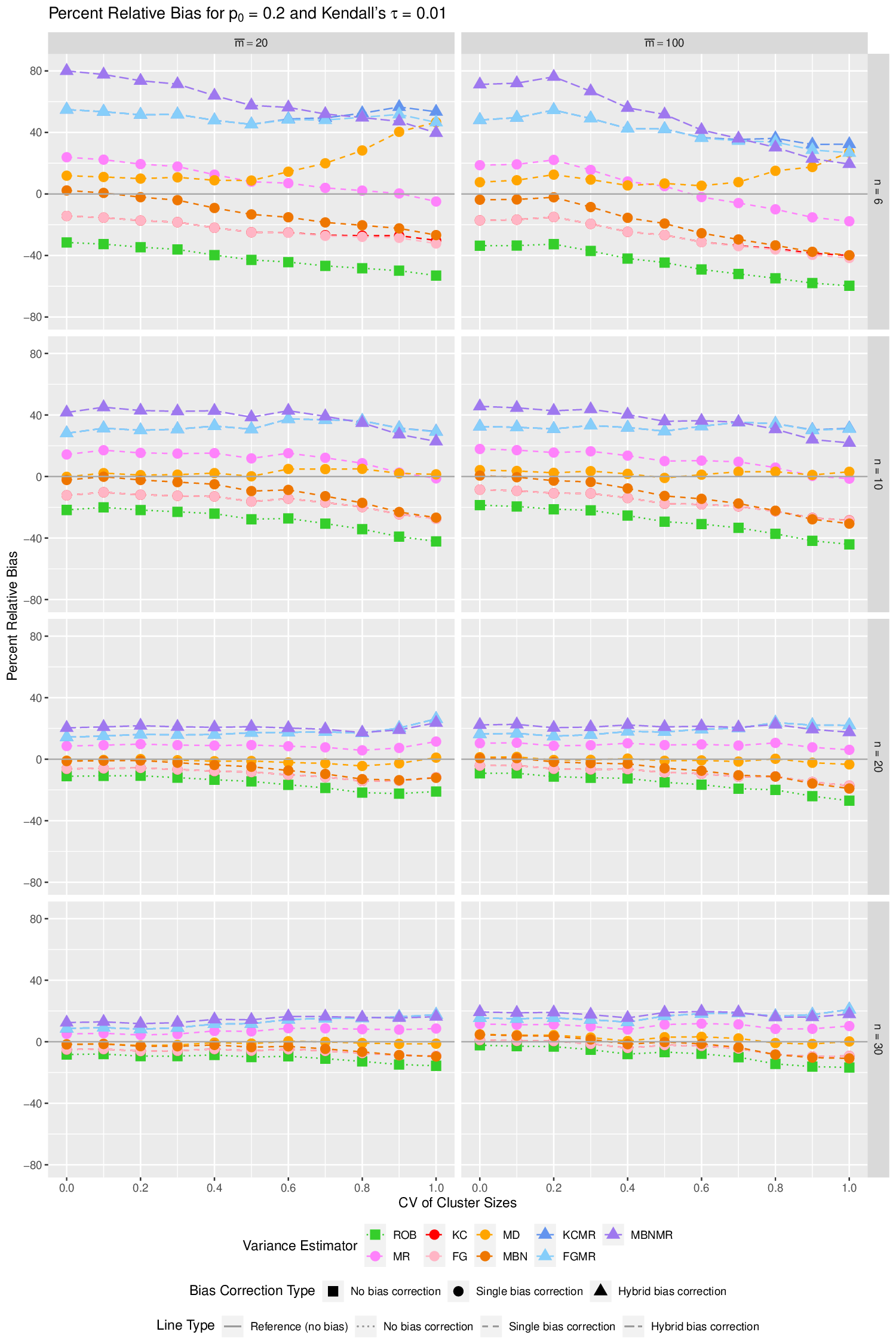}
\caption{Percent relative bias of different variance estimators for $p_0 = 0.2$ and $\tau = 0.01$ under the marginal Cox model.}
\label{fig:b001}
\end{figure}

\begin{figure}[htbp]
\centering
\includegraphics[scale=0.51]{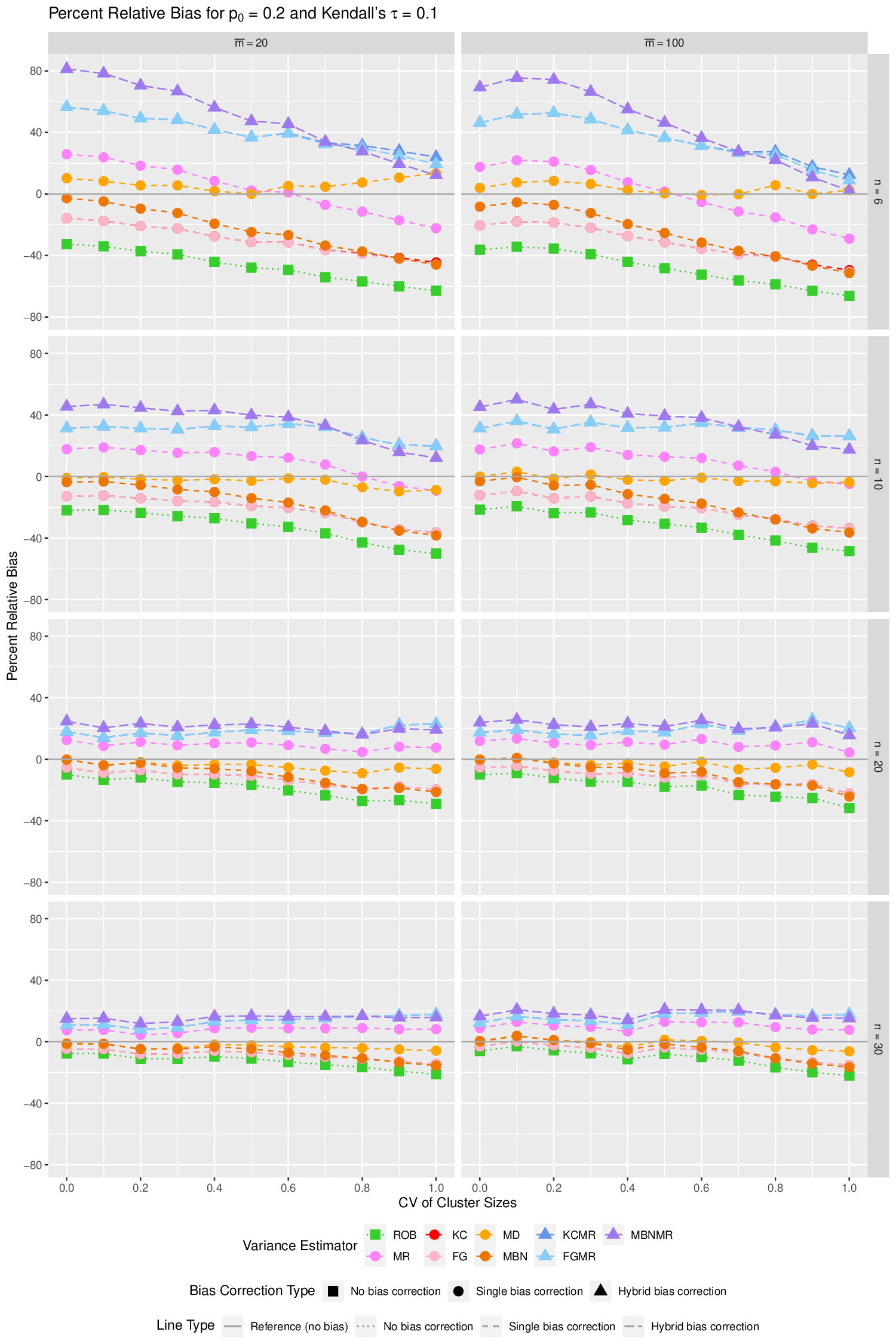}
\caption{Percent relative bias of different variance estimators for $p_0 = 0.2$ and $\tau = 0.1$ under the marginal Cox model.}
\label{fig:b01}
\end{figure}

Figure \ref{fig:b001} shows the results for the percent relative bias of different variance estimators, for $p_0 = 0.2$, $\tau = 0.01$ and $\overline{m} = 20 \text{ or } 100$, with gray lines indicating no bias. To better visualize the comparison, we excluded the MDMR estimator since it tended to substantially overestimate $\text{Var}(\widehat{\beta})$, especially for $n = 6$. As expected, the ROB estimator severely underestimated $\text{Var}(\widehat{\beta})$, especially when the number of clusters did not exceed $20$. Among the bias-corrected sandwich variance estimators, the MD estimator performed well with the smallest bias for $n \geq 10$, but led to some positive bias for $n = 6$ and CV $\geq 0.5$ (the most challenging scenarios). In general, the MR estimator tended to over-correct the variance with positive bias, but can exhibit negative bias under $n = 6$ and CV $\geq 0.6$; the KC, FG, and MBN estimators tended to show negative bias, especially when CV $\geq 0.5$, which is different from the findings in \citet{Lu2007} for binary outcomes. The three hybrid bias corrections, KCMR, FGMR, and MBNMR estimators, tended to over-correct the variance with positive bias. The similar pattern was observed in Figure \ref{fig:b01}, for $p_0 = 0.2$, $\tau = 0.1$, $\overline{m} = 20 \text{ or } 100$. In fact, all other combinations of $\tau$ and $\overline{m}$ led to this same pattern; the full results for the percent relative bias of all variance estimators under $p_0 = 0.2$ are presented in Web Figures 1-10.

\begin{figure}[htbp]
\centering
\includegraphics[scale=0.51]{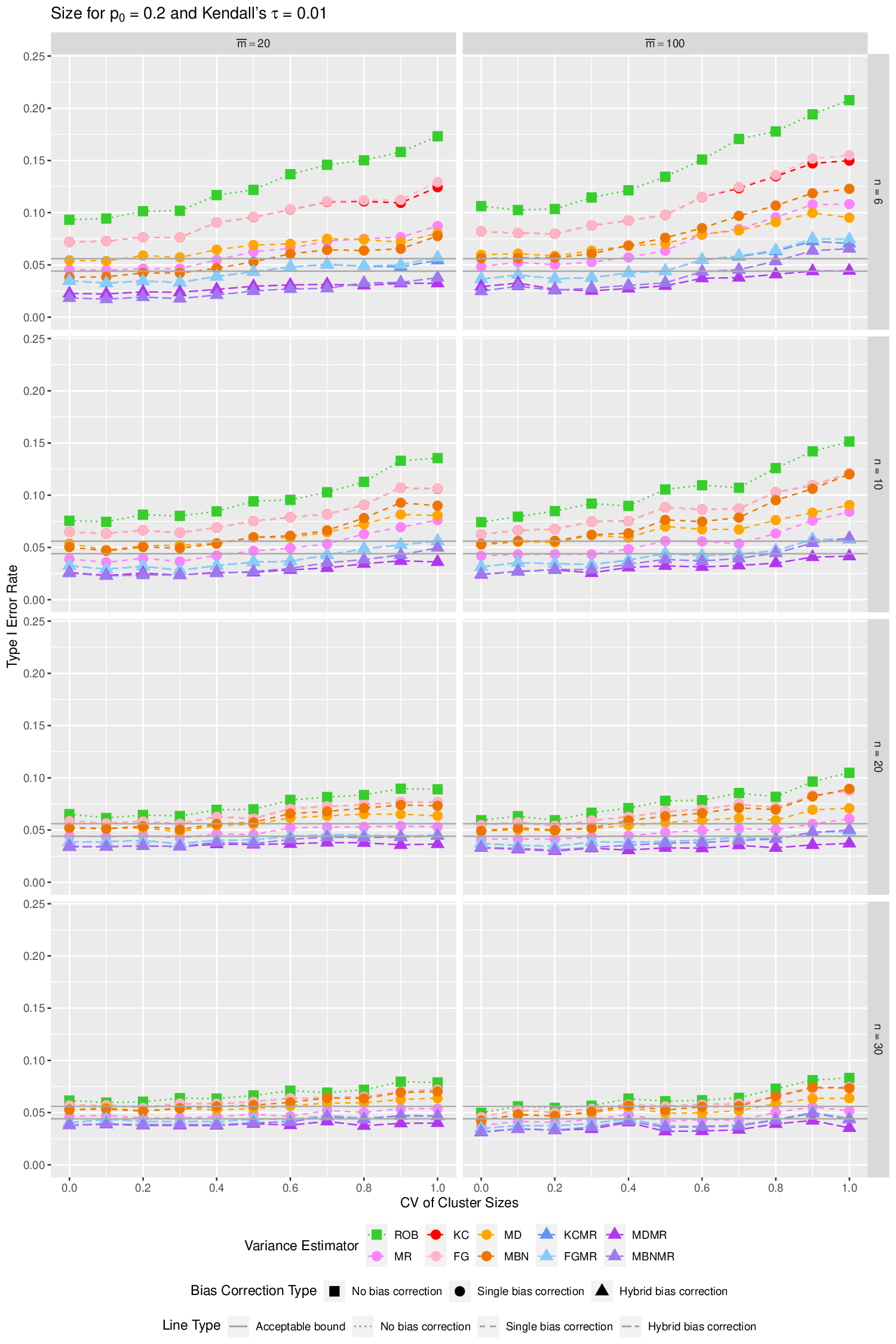}
\caption{Empirical type I error rates of intervention effect tests for $p_0 = 0.2$ and $\tau = 0.01$ under the marginal Cox model, based on different variance estimators.}
\label{fig:s001}
\end{figure}

\begin{figure}[htbp]
\centering
\includegraphics[scale=0.51]{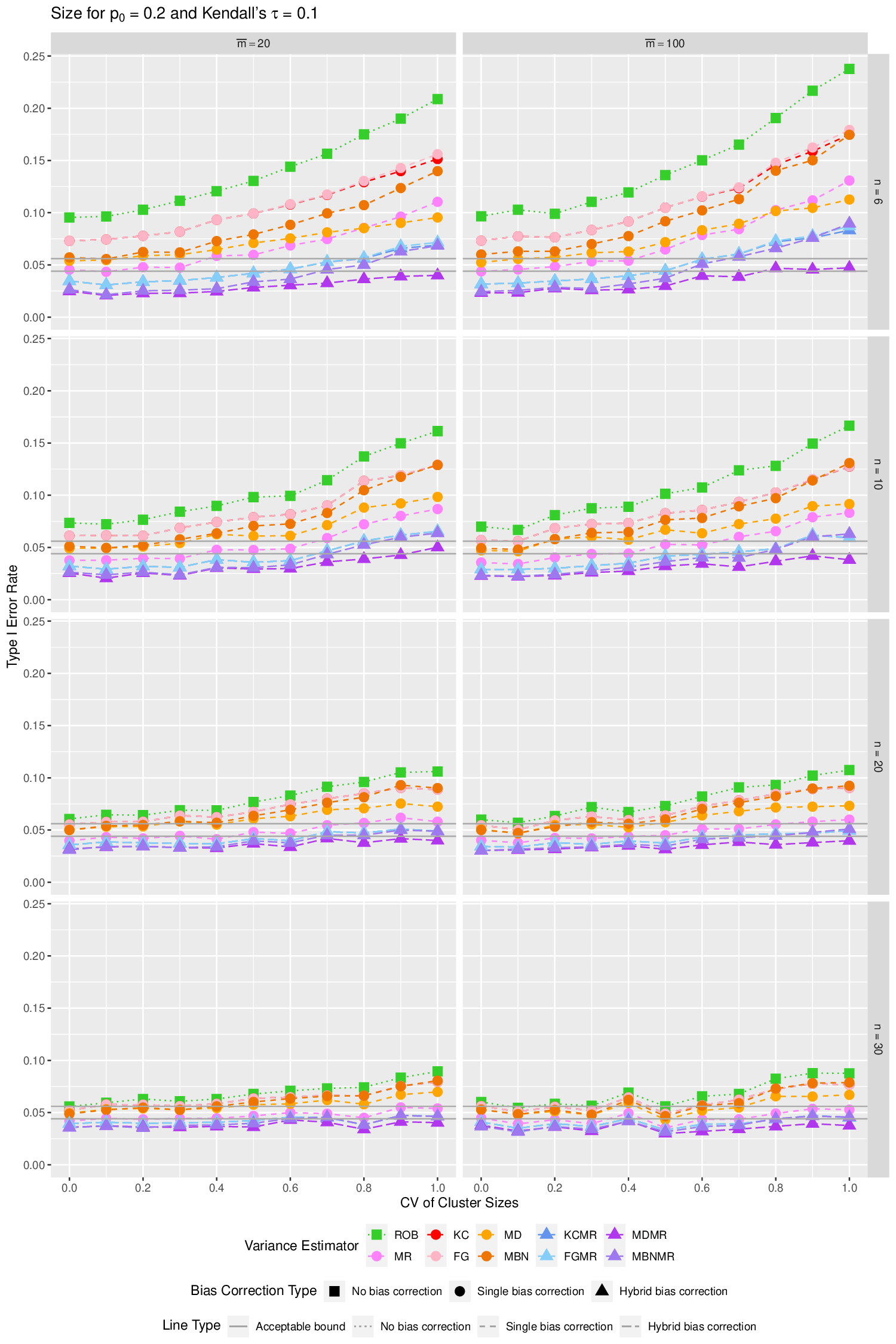}
\caption{Empirical type I error rates of intervention effect tests for $p_0 = 0.2$ and $\tau = 0.1$ under the marginal Cox model, based on different variance estimators.}
\label{fig:s01}
\end{figure}

Turning to hypothesis testing, Figure \ref{fig:s001} summarized the results for empirical type I error rates for $t$-tests using different variance estimators, with the acceptable bounds labeled in gray lines. From Figure \ref{fig:s001}, the test coupled with the ROB estimator had the worst performance with inflated type I error rates throughout, with the worst result being over $20\%$; all bias corrections improved the type I error rate results. In particular, the test with the MD estimator led to acceptable type I error rates only for CV $\leq 0.4$, but provided inflated type I error rates otherwise; tests with the KC or FG estimator often gave inflated type I error rates, whereas the tests with the MR and MBN estimator led to type I error results from acceptable to inflated as CV increased. Among the hybrid bias corrections, tests with the KCMR and FGMR estimators led to very similar results with KCMR marginally better (from slightly conservative to acceptable as CV increased), \textcolor{black}{except for a few extremely challenging scenarios under $n = 6$ and CV $\geq 0.8$, where the tests became liberal}; in comparison, the tests with the MDMR and MBNMR estimators tended to provide more conservative results. This general pattern was also observed with $\tau = 0.1$ and $\overline{m} = 20 \text{ or } 100$, as shown in Figure \ref{fig:s01}, as well as all other combinations of $\tau$ and $\overline{m}$, for $p_0 = 0.2$. The full results for type I error rates based on different variance estimators for $p_0 = 0.2$ are presented in Web Figures 11-20.

Overall, when CV $\leq 0.4$, the MD bias-corrected variance estimator (and the associated $t$-test) performed best considering both the percent relative bias and empirical type I error rates; when CV $\geq 0.5$, the $t$-test with the KCMR bias correction performed best with respect to the test size, with exceptions under $n = 6$ and CV $\geq 0.8$, in which case no bias corrections could lead to close to nominal test size. The contradiction between the positive bias of the KCMR bias-corrected sandwich variance estimator and the satisfactory test size of the associated $t$-test is the result of the variability of the KCMR sandwich variance estimator when CV of cluster sizes increases. Specifically, the tendency for over-rejecting the null due to the variability of the KCMR sandwich estimator happens to offset the positive bias and thus give the nominal test size. A similar seemingly contradictory result was observed in \citet{Lu2007} when conducting GEE inference of binary outcomes, but was based on the normality-based test with a different member of the bias-corrected variance estimator (the MD estimator). 

The results for $p_0 = 0.5$ had very similar patterns to those for $p_0 = 0.2$, and are presented in Web Figures 21-30 and 31-40 for the percent relative bias and empirical type I error rates, respectively.

\subsection{\textcolor{black}{Additional simulation scenarios}}\label{sec:add}
\textcolor{black}{
In addition to the above main simulation study, we also expanded the scenarios from different aspects to further enlarge the scope for practical recommendations. For simplicity, we only considered $p_0 = 0.2$ (administrative censoring only) for the expanded simulation study.}

\begin{enumerate} 
\itemsep0em 
\item[(i)]
\textcolor{black}{
Although the main numerical study focused on a CRT with only one binary cluster-level intervention indicator (often used as a primary analysis), cluster-varying or cluster-constant covariates might be considered in a secondary adjusted analysis or subgroup analysis in practice. Therefore, we additionally considered a two-arm CRT with two covariates in model \eqref{eq:ph} for an expanded simulation study: a binary cluster-level intervention indicator and an additional individual-level or cluster-level covariate, generated from the standard normal distribution. We chose $p_0 = 0.2$, $\tau = 0.01$, $\overline{m} = 20 \text{ or } 100$, and kept other parameters the same as in Section \ref{sec:msd}.
}

\textcolor{black}{
With an additional individual-level covariate, Web Figures 41 and 42 summarize the percent relative bias and empirical type I error rates for the $t$-tests, respectively, using different variance estimators. We observed that the patterns were very similar to those in Figures \ref{fig:b001} and \ref{fig:s001}, respectively, indicating that adding an additional individual-level covariate did not substantially affect the performances of bias-corrected variance estimators.
}

\textcolor{black}{
With an additional cluster-level covariate, Web Figures 43 and 44 present the percent relative bias and empirical type I error rates based on the $t$-tests, respectively, using different variance estimators. With exceptions under $n = 6$ and $10$ for the percent relative bias, the patterns were similar to those in Figures \ref{fig:b001} and \ref{fig:s001}, respectively. When $n = 10$, both of the MD and KCMR estimators had high percent relative bias, but the MD estimator provided acceptable type I error rates when CV $\leq 0.4$, and the KCMR estimator provided acceptable type I error rates when $0.5 \leq$ CV $\leq 0.7$, which is similar to the findings in Section \ref{sec:sr}. No bias corrections could lead to acceptable test size under $n = 6$, or under $n = 10$ and CV $\geq 0.8$. When the cluster size is highly variable, including more than $10$ clusters in CRTs would be recommended when adding an additional cluster-level covariate in the analysis. 
}

\item[(ii)]
\textcolor{black}{
While we included a range of values for the CV of cluster sizes (from $0$ to $1$) covering most situations in practice, it is possible that a CRT could have a larger variation of cluster sizes. Therefore, we conducted simulations with $p_0 = 0.2$, $\tau = 0.01$, the CV of cluster sizes ranging from $1.1$ to $1.5$ by an increment of $0.1$, and other parameters same as in Section \ref{sec:msd}.
}

\textcolor{black}{
Web Figures 45 and 46 summarize the percent relative bias and empirical type I error rates for the $t$-tests, respectively, using different variance estimators under larger CV of cluster sizes. The patterns were similar to those under CV $\geq 0.5$ in Figures \ref{fig:b001} and \ref{fig:s001}, respectively, confirming that the KCMR bias-corrected sandwich variance estimator performed best when CV $\geq 0.5$, including when $1< \text{CV}\leq 1.5$.
}

\item[(iii)]
\textcolor{black}{
Even though this article focused on the type I error rate for hypothesis testing, we extended our simulations to investigate whether the performance of the proposed bias-corrected sandwich variance estimators changed under a non-null marginal hazard ratio (non-null intervention effect), as measured by the empirical coverage probability of the associated $95\%$ confidence interval (CI). Specifically, keeping other parameters the same as in Section \ref{sec:msd}, we chose $p_0 = 0.2$, and the number of clusters $n$ ranging from $20$ to $30$ across all scenarios. At a $95\%$ nominal level, we considered an empirical coverage probability from $94.4\%$ to $95.6\%$ as acceptable by the margin of error from a binomial model with $5000$ replications \citep{morris2019using}.
}

\textcolor{black}{
Web Figure 47 shows the results for empirical coverage probabilities of different variance estimators under $p_0 = 0.2$, with gray lines indicating the acceptable bounds. As expected, the ROB estimator always led to under-coverage, especially when the CV of cluster sizes were large. Among the single bias-corrected sandwich variance estimators, the MD and MBN estimators performed well for CV $\leq 0.4$, but gave under-coverage for CV $\geq 0.5$; the KC and FG estimators often led to under-coverage. In general, the MR estimator was not stable, providing the results from over-coverage to under-coverage as the CV of cluster sizes increased. Among the hybrid bias corrections, the KCMR, FGMR, and MBNMR estimators tended to provide the results from over-coverage to acceptable as the CV of cluster sizes increased; the MDMR estimator always led to over-coverage. Overall, these results corroborated our choice of the MD bias-corrected variance estimator for CV $\leq 0.4$ and the KCMR bias-corrected sandwich variance estimator for CV $\geq 0.5$.
}
\end{enumerate}

\subsection{Practical recommendations}\label{sec:prac}
The challenge we face in the simulation study led us to suggest that practitioners include at least $10$ clusters in CRTs with right-censored, time-to-event outcomes, when the cluster sizes are anticipated to be variable, \textcolor{black}{based on the desire to generate a valid test for both unadjusted analysis and adjusted analysis with an additional individual-level covariate. In the meantime, when the cluster size is expected to be highly variable (e.g., $\text{CV}\geq 0.8$), including more than $10$ clusters would be recommended for adjusted analysis with an additional cluster-level covariate}. Otherwise, when the number of clusters is small (e.g. $n = 6$) and the CV of cluster sizes is large (CV $\geq 0.8$), there is not a convenient way to modify the robust sandwich variance estimator to maintain the test size under the marginal Cox model; this recommendation is generally in line with recommended practices in CRTs because it may be more unlikely for the study to provide sufficient power with fewer than $10$ clusters in the absence of an overwhelming effect size. In general, the MD bias-corrected sandwich variance estimator has the smallest bias throughout. When testing the intervention effect in CRTs with a small number of clusters, our results suggest that, for Wald $t$-tests, the use of the MD bias-corrected sandwich variance estimator can maintain the nominal test size \textcolor{black}{and provide adequate empirical coverage probabilities}, and is robust to the moderate variation of cluster sizes (CV $\leq 0.4$). However, under larger variations of cluster sizes, the KCMR bias-corrected sandwich variance estimator should be used instead to maintain the nominal test size \textcolor{black}{and provide adequate empirical coverage probabilities} due to the increased variability of all sandwich variance estimators. This observation that the optimal recommendation of the sandwich variance estimator depends on the CV of cluster sizes is consistent with \citet{li2015}, who recommended different bias-corrected variance estimators (the KC and FG variance estimators) with clustered binary outcomes depending on the cluster size variation. Our recommendation for clustered time-to-event outcomes complements \citet{li2015} and serves to refine our understanding of best practices for analyzing CRTs.

\section{Application to the STOP-CRC Cluster Randomized Trial}\label{sec:app}
We analyze outcome data from the STOP CRC trial, a two-arm parallel CRT described in Section \ref{sec:intro}. The STOP CRC trial compared two strategies - an EHR-embedded program and usual care - for colorectal cancer screening rates within 12 months of enrollment \citep{coronado2018effectiveness}. We consider the outcome of interest as the time to completion of colorectal cancer screening (returning the FIT kits test result), which was administratively censored at 12 months. Randomization was conducted at the level of health center clinic (cluster) with equal allocation to the two arms. We consider the subgroup of nonwhite participants, to illustrate the scenario with a high CV of cluster sizes and to serve as a scenario where the application of bias-corrected sandwich variance estimator can lead to a different conclusion from the standard analysis. There were $4543$ nonwhite participants in $26$ clusters, of which $2513 \ (55.32\%)$ were female. The mean (standard deviation [SD]) age of all $4543$ nonwhite participants was $58.72 \ (6.51)$ years. The number of nonwhite participants in the $26$ clusters ranged from $8$ to $1054$, with mean (SD) of $174.73 \ (246.95)$. Consequently, a notable feature of this CRT was the small number of clusters ($n=26$) with considerable cluster size variability (CV = $1.41$).

We consider a proportional hazards model \eqref{eq:ph} including only a binary cluster-level intervention indicator  $Z_\ii$ ($p=1$), and performed the comparison among the ROB estimator and $9$ bias-corrected sandwich variance estimators. Table \ref{res:app} presents the estimated intervention effect (in population-averaged hazard ratio) as well as the Wald $t$-test result based on different variance estimators. Evidently, the $95\%$ confidence intervals with bias-corrected variance estimators were wider than that with the ROB estimator. Because the CV of cluster sizes is large and the number of clusters ($n=26$) is between $20$ and $30$, the $t$-test with the KCMR bias-corrected sandwich variance estimator may have close to nominal size. In this case, we could conclude that the hazard ratio due to intervention is $2.102$ ($95\%$ CI: $[0.780, 5.191]$, $p$-value = $0.141$) in the nonwhite subgroup; since the $p$-value exceeds $0.05$, we fail to reject the null of non-intervention effect. It should be noted that the KCMR and FGMR estimators provided the same results in this application, due to inclusion of only the intervention effect indicator in the Cox model. Finally, if one proceeds with the $t$-test with the ROB variance estimator in this analysis, the resulting $p$-value is smaller than $0.05$ and one may reach an over-confident conclusion by rejecting the null, since we know the ROB estimator is biased towards zero with fewer than $30$ clusters. 

\begin{table}[htbp]
\caption{Analysis results of the STOP CRC data (for the sub-population of nonwhite).}\label{res:app}
\centering
%\resizebox{\textwidth}{!}{%
\begin{threeparttable}
\begin{tabular}{lccc}
\toprule
Variance Estimator & Log of Hazard Ratio\tnote{a} ($95\%$ CI\tnote{b} ) & Hazard Ratio ($95\%$ CI\tnote{b} ) & $p$-value \\\midrule
%MB & 0.699 (0.544, 0.854) & 2.012 (1.723, 2.350) & 0.000 \\
ROB & 0.699 (0.151, 1.248) & 2.012 (1.162, 3.483) & 0.015 \\
MR & 0.699 (-0.011, 1.410) & 2.012 (0.989, 4.094) & 0.053 \\
KC & 0.699 (-0.018, 1.416) & 2.012 (0.983, 4.120) & 0.055 \\
FG & 0.699 (-0.018, 1.416) & 2.012 (0.983, 4.120) & 0.055 \\
MD & 0.699 (-0.274, 1.672) & 2.012 (0.761, 5.323) & 0.151 \\
MBN & 0.699 (0.129, 1.270) & 2.012 (1.137, 3.560) & 0.018 \\
KCMR & 0.699 (-0.249, 1.647) & 2.012 (0.780, 5.191) & 0.141 \\
FGMR & 0.699 (-0.249, 1.647) & 2.012 (0.780, 5.191) & 0.141 \\
MDMR & 0.699 (-0.605, 2.004) & 2.012 (0.546, 7.417) & 0.280 \\
MBNMR & 0.699 (-0.040, 1.438) & 2.012 (0.961, 4.212) & 0.063 \\
\bottomrule
\end{tabular}\smallskip
\begin{tablenotes}\small
\item[a] Estimate of $\beta$ in model \eqref{eq:ph} with only one covariate of a binary cluster-level intervention indicator.\smallskip
\item[b] CI: Confidence interval from the Wald $t$-test with $n-1$ degrees of freedom.%\smallskip
\end{tablenotes}
\end{threeparttable}
%}
\end{table}

\section{Discussion}\label{sec:dis}
In this article, we propose $9$ bias-corrected sandwich variance estimators for CRTs with time-to-event data analyzed through the marginal Cox model, to address the negative bias in the standard robust sandwich variance estimator with a limited number of clusters (a prevailing practice in CRTs). We also conduct a comprehensive simulation study to evaluate the small-sample properties of the variance estimators as well as the associated Wald $t$-tests to maintain appropriate type I error rates \textcolor{black}{and provide adequate empirical coverage probabilities}. Above all, the uncorrected sandwich variance estimator could lead to substantially inflated type I error rates, sometimes even larger than $20\%$. \textcolor{black}{Our simulation results suggest that the Wald $t$-test coupled with a bias-corrected sandwich variance estimator can maintain the nominal test size, provide adequate empirical coverage probabilities, and generate reliable inferences, for as few as $10$ clusters in both unadjusted analysis and adjusted analysis with an additional individual-level covariate; the choice of bias-corrected sandwich variance estimators should take the variation of cluster sizes into account, as we have summarized in Section \ref{sec:prac}. For adjusted analysis with an additional cluster-level covariate, the chosen bias-corrected sandwich variance estimators also work well for as few as $10$ clusters when the cluster size variation is moderate; however, when the cluster size is highly variable ($\text{CV}\geq 0.8$), including more than $10$ clusters would be recommended.} When the number of clusters is extremely small (say, $n = 6$), the marginal Cox model should be used carefully, as the size of Wald $t$-tests associated with a bias-corrected sandwich variance estimator could only maintain nominal size when the CV of cluster sizes does not exceed $0.8$. In any case, the standard sandwich variance estimator should be avoided in practice when the number of clusters is small ($n \leq 30$), which may lead to over-confident results as in our real data application in Section \ref{sec:app}. On the contrary, there is no harm asymptotically in applying the bias-corrected sandwich variance estimators for inference. To facilitate the implementation of the proposed bias-corrected sandwich variance estimators, we have also developed an R package \textbf{CoxBcv}.

We have considered the marginal Cox regression with the working independence correlation structure as the basis of our work. This approach is currently the standard implementation in existing software packages, but is subject to potential limitations. For example, the measure of intracluster correlation is often of substantial interest in CRTs, for which purpose a second-order estimating equations based on the martingale residuals can be constructed following \citet{cai1995estimating}. The basic principles discussed in our article should still be directly applicable to the marginal Cox analysis assuming a non-independent working correlation structure. However, the optimal bias-corrections warrants future research. While this more complex formulation of paired estimating equations can be of substantial interest, we are currently not aware of any statistical packages that implement this approach, \textcolor{black}{perhaps due to the associated computational challenges when the cluster sizes are relatively large (which is not uncommon in CRTs)}. In future work with the paired estimating equations, an additional useful direction is to develop bias-corrections for the correlation estimating equations, similar to the matrix-adjusted estimating equations proposed by \citet{preisser2008}. Finally, we acknowledge that our simulation study is not exhaustive and has so far focused on the unadjusted analysis without additional covariates, and adjusted analysis with an additional covariate beyond the cluster-level intervention variable. 
We anticipate future work to generalize our recommendations for other study designs with clustered time-to-event data and more complex covariate adjustment patterns. The availability of our R package can facilitate the design and execution of future simulation studies, perhaps motivated by scenarios where adjusting for multiple covariates is necessary to generate valid inference.

%\backmatter

\section*{Acknowledgements}
This work is partially supported within the NIH Health Care Systems Research Collaboratory by the NIH Common Fund through cooperative agreement U24AT009676 from the Office of Strategic Coordination within the Office of the NIH Director. This work is also supported by the NIH through the NIH HEAL Initiative under award number U24AT010961, and by awards U01OD033247 and R01DC020026 from the NIH. The authors thank Dr. Gloria Coronado from the Kaiser Permanente Center for Health Research for providing permission for us to use deidentified data from the STOP CRC study. Funding for the STOP CRC study was provided by awards from the NIH (UH2AT007782 and 4UH3CA18864002). The content of the work presented is solely the responsibility of the authors and does not necessarily represent the official views of the NIH or its HEAL Initiative. The authors thank Mr. Can Meng for computational assistance with the simulation study reported in Section 4.3. The authors are also grateful to the Associate Editor and two anonymous reviewers for their constructive comments and suggestions, which have improved the exposition of this work.

\bibliographystyle{biom}
\bibliography{SCRTbib}

\section*{Supporting Information}
Web Appendices and Figures referenced in Sections \ref{sec:power}-\ref{sec:sim} are available on GitHub at \url{https://github.com/XueqiWang/CoxBcv_SI}. The R package \textbf{CoxBcv} is openly available on the Comprehensive R Archive Network (CRAN) and on GitHub at \url{https://github.com/XueqiWang/CoxBcv_R_package}.

\label{lastpage}
\end{document}